\documentclass[aps,preprint,pra]{revtex4}
\usepackage{graphicx}

\newcommand{\x}{{\bf x}}
\newcommand{\xt}{(\x,t)}

\begin{document}

\title{Protective measurements and the PBR theorem}\thanks{In {\it Protective Measurement:  Towards a New Understanding of Quantum Mechanics}, ed. Shan Gao, to appear.}

\author{Guy Hetzroni}
\affiliation{Program in the History and Philosophy of Science, The Hebrew University of Jerusalem, Jerusalem 91905, Israel}

\author{Daniel Rohrlich}
\affiliation{Department of Physics, Ben Gurion University of the Negev, Beersheba
84105 Israel}

\date{\today}

\begin{abstract}
Protective measurements illustrate how Yakir Aharonov's fundamental insights into quantum theory yield new experimental paradigms that allow us to test quantum mechanics in ways that were not possible before.  As for quantum theory itself, protective measurements demonstrate that a quantum state describes a single system, not only an ensemble of systems, and reveal a rich ontology in the quantum state of a single system.  We discuss in what sense protective measurements anticipate the theorem of Pusey, Barrett, and Rudolph (PBR), stating that, if quantum predictions are correct, then two distinct quantum states cannot represent the same physical reality.
\end{abstract}

\pacs{}

\maketitle

Although protective measurements \cite{av,aav} are a new tool for quantum theory and experiment, they have yet to find their way into the laboratory; also theorists have not put them to best use, beyond a 1993 paper by Anandan on ``Protective measurements and quantum reality"  \cite{jeeva}.  Below, in Sect. \ref{sectionProtective}, we point out that protective measurements offer  new experimental tests of quantum mechanics, and we review recent experiments attempting to measure quantum wave functions.  In Sect. \ref{sectionPBR}, we present the Pusey-Barrett-Rudolph (PBR) theorem and discuss their conclusion that the quantum state represents physical reality, and in Sect. \ref{sectionProtectivePBR}, we discuss in what sense protective measurements anticipate this conclusion.

\section{Protective measurements: implications for experiment and theory}
\label{sectionProtective}

In 1926, Schr\"odinger postulated his equation for ``material waves" in analogy with light waves: paths of material particles---which obey the principle of least action---are an approximation to material waves, just as rays of light---which obey the principle of least time---are an approximation to light waves \cite{material}. But Born soon discarded ``the physical pictures of Schr\"odinger," \cite{born} and gave the ``material wave" $\Psi\xt$ a new interpretation:  $\vert\Psi \xt \vert^2$ is the probability density to find a particle at $\x$ at time $t$.  Even Schr\"odinger was obliged to accept Born's interpretation.  But Born's interpretation limits the correspondence between quantum theory and experiment, in the following sense:  for a {\it single} particle, $\Psi \xt$ seems {\it not} to be measurable; to measure a probability density, we need to prepare $\Psi \xt$ on an ensemble.  Thus, part of what quantum theory describes---the wave function $\Psi \xt$ of a single particle---does not correspond to anything experiments can measure.  The paradigm of protective measurements \cite{av, aav, QP}, by contrast, makes the correspondence explicit:  experiments {\it can} measure the wave function of a single particle!  Protective measurements make it possible to measure the expectation value of any operator $A$ in any state $\vert \Psi \rangle$ using a single system prepared in the state $\vert\Psi\rangle$, and thus to reconstruct $\vert \Psi \rangle$.  An ensemble of identical systems in the state $\vert\Psi \rangle$ is not necessary.  By the same token, the method of protective measurement allows us to test quantum mechanics in ways that were never considered before, {i.e.} to verify expectation values measured on an isolated system.

A recent experiment of Lundeen {et al.} \cite{jsl} measured the transverse spatial wave function of a photon propagating as a plane wave.  These authors do not mention ``protective measurement"---they refer only to ``weak measurement"---and their experiment differed from a protective measurement in two ways.  First, the measurement was applied to an ensemble of photons rather than to a single photon; second, what they measured was not an expectation value but the weak value \cite{weak} of the projection operator $\Pi_x \equiv\vert x\rangle\langle x\vert$ onto a transverse position $x$:
\begin{equation}
\left\langle \Pi_x\right\rangle_w = {{\langle p \vert x\rangle \langle x \vert\Psi\rangle}\over{\langle p\vert \Psi\rangle}}=  {{e^{-ipx/\hbar} \Psi (x)}\over{\langle p\vert \Psi\rangle}}~~~,
\end{equation}
where $\Psi(x)$ is the (preselected) transverse wave function to be measured, $\vert p\rangle$ is a (postselected) transverse momentum eigenstate with momentum $p$, and the postselected momentum is $p=0$.  Then the weak value is proportional to $\Psi (x)$ and the initial wave function (both real and imaginary parts) are measured as a function of $x$.

Although the weak measurement of Lundeen {et al.} is not a protective measurement, protective measurements are a form of weak measurements \cite{apc}.  If the pre- and postselected states $\vert \Psi_{in}\rangle$, $\vert \Psi_{fin}\rangle$ of a weak measurement of $A$ are the same, the measured weak value $\langle A \rangle_w$ is the expectation value of $A$ in the state $\vert \Psi_{in}\rangle$:
\begin{equation}
\langle A\rangle_w ={{\langle \Psi\vert A \vert \Psi\rangle}\over {\langle \Psi \vert\Psi \rangle}} = \langle A \rangle~~~,
\end{equation}
where $\vert \Psi_{fin}\rangle =\vert \Psi\rangle = \vert \Psi_{in}\rangle$. In a typical weak measurement, the pointer of a measuring device is coupled to an $ensemble$ of systems pre- and postselected in the state $\vert \Psi\rangle$, and shifts as the expectation value $\langle A \rangle$ accumulates from all the systems in the ensemble.  (Note that, while the postselection in many weak measurements is improbable, here the postselection is the most probable.)  By contrast, a protective measurement is essentially a weak measurement repeated on the {\it same} system, and the pointer shifts as the expectation value $\langle A \rangle$ accumulates from the repeated measurement.  Repeated post- and preselections insure the protection, and most-probable postselections insure the adiabaticity.  In effect, a repeated measurement of $A$ on a single system in the state $\vert \Psi\rangle$ yields the same result as single measurements of $A$ on an ensemble of systems pre- and postselected in the state $\vert\Psi\rangle$; however, only the first kind of measurement---protective measurement---explicitly manifests the quantum state of a {\it single} system.

A more recent experiment by Stodolna {et al.} \cite{stodolna} maps the nodal structure the $n=30$ Rydberg level of hydrogen in a uniform electric field. (See also a related experiment by Cohen {et al.} \cite{cohen}, and measurement of molecular wave functions by L\"uftner {et al.} \cite{lu}.  These three papers, as well, do not mention protective measurements.)  An electron excited to this level is quasibound:  classically, it cannot escape, but it can escape via quantum tunnelling and then accelerates in the electric field towards a phosphor screen and CCD camera.  The measurement is not adiabatic.  The electrons, released from an initial beam of hydrogen atoms via photoionization, image the Rydberg wave function on the screen.  So, although the experiment involves an ensemble of atoms, each atom contributes independently to the measurement, i.e. reveals a different feature of the initial wave function.

Alongside the experimental development, protective measurements allow us to develop new intuitions for quantum theory.  They demonstrate that the members of any ensemble of systems prepared in a given quantum state have much more in common than what previous measurements showed.  Not only do all systems prepared in an eigenstate of an operator $A$ yield the same eigenvalue when subjected to a measurement of $A$; they yield the same expectation values for {\it any} operator that can be measured on the system.  Thus, an ensemble of systems prepared in a given state share a ``group identity" which is much richer than a shared eigenvalue:  it includes every expectation value that can be measured on the state.  The next two sections show that this group identity has implications for the ontology of the quantum state.

\section{The Pusey-Barrett-Rudolph (PBR) theorem}
\label{sectionPBR}

Probability distributions are often interpreted as subjective, {i.e.} as representing an observer's knowledge (or ignorance) about a system. Is this also the correct way to interpret probabilities derived from quantum states? There seem to be good reasons to favor such an interpretation, because the alternative interpretation---that the quantum state is no more than a description of the reality of a system---is disturbing in several respects.  As a description of reality, the quantum state apparently exhibits instantaneous collapse over unbounded spatial regions. It also superposes properties that are (classically) mutually exclusive. Entanglement implies that the quantum state of a composite system cannot be reduced to states of the component systems. These peculiarities are less troubling if the quantum state represents information about a system, rather than the system's actual physical state \cite{heisenberg,hartle}.

Harrigan and Spekkens \cite{HS} gave this question a precise formulation. If the quantum state is a representation of knowledge about an unknown and possibly inaccessible physical state, it does not depend solely on the properties of the system. It depends also on the information available to the observer. Therefore, if the quantum state represents subjective knowledge, at least some {\it physical} state has to be compatible with more than one {\it quantum} state. The probabilistic nature of the predictions of quantum theory seems to allow for such compatibility, as long as the two quantum states that can represent one reality are not orthogonal.

More formally, let $\lambda$ (which could be a number or a vector, and belongs to a space denoted $\Lambda$) be a complete specification of the physical state of a system, e.g. of an atom.  If a quantum state $\vert \Psi \rangle$ of that system corresponds to a single $\lambda$, then $\vert \Psi \rangle$ as well is a complete specification of the physical state of the system.  But in general, $\vert \Psi \rangle$ could correspond to a probability distribution $p_\Psi (\lambda)$ over the values of $\lambda$.  If so, then the values of $\lambda$ play the role of hidden variables of a system in the state $\vert \Psi\rangle$. Now consider two possible states of the system, $\vert \Psi\rangle$ and $\vert \Phi\rangle$. If $\vert \Psi\rangle$ and $\vert\Phi\rangle$ are orthogonal, then their respective probability distributions, $p_\Psi (\lambda)$ and $p_\Phi (\lambda)$, must be {\it non-overlapping}, {i.e.} $p_\Psi (\lambda)p_\Phi (\lambda)=0$ for all $\lambda$. Otherwise, a prediction of quantum theory---namely, that a measurement of the projection operator $\Pi_\Phi \equiv \vert \Phi\rangle\langle \Phi\vert$ on a system prepared in the orthogonal state $\vert \Psi\rangle$ yields 0---will fail (since measuring devices respond only to the physical state).  But if $\vert \Psi\rangle$ and $\vert\Phi\rangle$ are not orthogonal (and not identical), it is conceivable that $\vert \Psi\rangle$ and $\vert\Phi\rangle$ could overlap.  If $\vert \Psi\rangle$ and $\vert\Phi\rangle$ overlap, so that for some $\bar{\lambda}$ we have $p_\Psi (\bar{\lambda})p_\Phi (\bar{\lambda})\ne 0$, then the two distinct quantum states $\vert \Psi\rangle$ and $\vert\Phi\rangle$ could represent the same physical reality $\bar{\lambda}$.  Conversely, if $p_\Psi (\lambda)p_\Phi (\lambda)=0$ for all $\lambda$ and for any two distinct states $\vert \Psi\rangle$ and $\vert\Phi\rangle$, then quantum states represent physical reality.

What is beautiful about this formulation is that it cleanly pulls apart two different questions about the quantum state.  The first question---the title of the famous EPR paper \cite{EPR} and of Bohr's reply \cite{bohr}---is whether the quantum state is a complete description of a physical state, {i.e.} whether one quantum state can represent more than one physical state.  (If a quantum state $\vert\Psi \rangle$ completely describes a physical state $\bar{\lambda}$, then $\vert\Psi \rangle$ cannot represent more than one physical state.)  The second question is whether two quantum states can represent one and the same same physical state.  The Pusey-Barrett-Rudolph (PBR) \cite{PBR} theorem states that if the predictions of quantum mechanics are correct, then the answer to the second question is negative, regardless of the answer to the first question:  no two distinct quantum states can represent the same physical reality, regardless of completeness.

The proof of the PBR theorem is technical.  Here we try to motivate the proof intuitively.  We begin by assuming that two nonorthogonal but distinct qubit states, $\vert 0\rangle$ and $\vert +\rangle$, with $\langle 0\vert +\rangle =1/\sqrt{2}$, represent in all cases exactly the same physical reality $\bar{\lambda}\in\Lambda$.  That is, both $p_0 (\lambda)$ and $p_+ (\lambda)$ vanish for $\lambda \ne {\bar{\lambda}}$.  Particles in a mixture of the states $\vert 0\rangle$ and $\vert +\rangle$ are fed into a device that measures an operator with nondegenerate eigenstates $\vert 0\rangle$ and $\vert 1\rangle$, where $\langle 0\vert 1\rangle =0$.  Quantum mechanics predicts that the device should sometimes find a particle in the state $\vert 1\rangle$, but only if the initial state was $\vert +\rangle$, never if the initial state was $\vert 0\rangle$.  But, by assumption, $\vert 0\rangle$ and $\vert +\rangle$ represent the same physical state $\bar{\lambda}$; hence there is {\it no way} the device can distinguish them, and, if it finds any particle in the state $\vert 1\rangle$, it must do so sometimes also when the particle's initial state was $\vert 0\rangle$.  Thus our assumption implies a violation of quantum predictions.

So far, the proof was easy because we assumed that $\vert 0\rangle$ and $\vert +\rangle$ can only represent a single physical reality $\bar{\lambda}$.  What if $\vert 0\rangle$ and $\vert +\rangle$ correspond to overlapping distributions $p_0 (\lambda)$ and $p_+ (\lambda)$?  Now the device could find particles in the state $\vert 1\rangle$ {\it only} for values of $\lambda$ for which $p_+ (\lambda)\ne 0$ and $p_0 (\lambda) =0$, {i.e.} not in the overlap of the distributions.  Hence the device need not violate quantum predictions:  it finds the state $\vert 1\rangle$ only when the initial state is not $\vert 0\rangle$.  To contradict quantum predictions, the device would have to measure an operator with a nondegenerate eigenstate $\vert -\rangle$ orthogonal to $\vert +\rangle$ as well as the nondegenerate eigenstate $\vert 1\rangle$ orthogonal to $\vert 0\rangle$.  Of course, no Hermitian operator can have $\vert 1\rangle$ and $\vert -\rangle$ as nondegenerate eigenstates.  What PBR showed, however, is that for a mixture of $pairs$ of particles prepared in the states $\vert 0\rangle \otimes\vert 0\rangle$, $\vert 0\rangle \otimes\vert +\rangle$, $\vert +\rangle \otimes\vert 0\rangle$ and $\vert +\rangle \otimes\vert +\rangle$, there is a nondegenerate operator, on the four-dimensional Hilbert space spanned by these eigenvectors, with the following property:  each of these four preparations is orthogonal to one of the operator's eigenstates.  Explicitly, the eigenstates are
\begin{eqnarray}\label{4vec}
\vert \xi_1\rangle &=& {1\over \sqrt{2}}\left( \vert 0\rangle \otimes\vert 1\rangle +\vert 1\rangle \otimes\vert 0\rangle\right)~~~,\cr
\vert \xi_2\rangle &=& {1\over \sqrt{2}}\left( \vert 0\rangle \otimes\vert -\rangle +\vert 1\rangle \otimes\vert +\rangle\right)~~~,\cr
\vert \xi_3\rangle &=& {1\over \sqrt{2}}\left( \vert +\rangle \otimes\vert 1\rangle +\vert -\rangle \otimes\vert 0\rangle\right)~~~,\cr
\vert \xi_4\rangle &=& {1\over \sqrt{2}}\left( \vert +\rangle \otimes\vert -\rangle +\vert -\rangle \otimes\vert +\rangle\right)~~~~.
\end{eqnarray}
Now the measuring device cannot avoid violating a prediction of quantum mechanics every now and then.  Note that the case of $\vert 0\rangle$ and $\vert +\rangle$ is special, because without the assumption $\langle 0\vert +\rangle =1/\sqrt{2}$ above, the states in Eq. (\ref{4vec}) would not be orthogonal.  The PBR proof of the general case is still more technical.

As an application of the PBR theorem, let us revisit the EPR paper \cite{EPR}.  Consider an entangled state
\begin{equation}
{1\over{\sqrt{2}}} \left[ \vert 0\rangle_A \otimes \vert 1\rangle_B -\vert 1\rangle_A \otimes \vert 0\rangle_B\right]
\end{equation}
of a particle pair shared by Alice and Bob, far apart in their respective laboratories.  If Alice measures $\vert 1\rangle_A {}_A\langle 1\vert - \vert 0\rangle_A {}_A\langle 0\vert$ on her particle, she might leave Bob's particle in the state $\vert 0\rangle_B$; but if she measures $\vert +\rangle_A {}_A\langle +\vert - \vert -\rangle_A {}_A\langle -\vert$ on her particle, she might leave Bob's particle in the state $\vert +\rangle_B$, which is distinct from $\vert 0\rangle_B$ and not orthogonal to it. The state $\vert 0\rangle_B$, claim EPR, must represent the same physical reality as the state $\vert +\rangle_B$, since no influence, including Alice's measurement, can propagate faster than the speed of light.  But according to the PBR theorem, if the predictions of quantum mechanics are correct, then the state $\vert 0\rangle_B$ {\it cannot} represent the same reality as $\vert +\rangle_B$.  We see that the EPR assumption---according to which Alice's measurement does not disturb Bob's particle---is incompatible with quantum mechanics. It is striking that both Bell's theorem \cite{bell, chsh} and the PBR theorem imply that EPR's demand for locality (Einstein separability) is incompatible with quantum mechanics, even though the PBR theorem does not mention locality.

\section{Protective measurements, PBR and the reality of $\vert \Psi \rangle$}
\label{sectionProtectivePBR}

Assuming that quantum predictions are correct, the PBR theorem implies that a quantum state representing an individual system also represents a part or all of the physical reality of that system.  Independently, protective measurements make it possible to measure expectation values, including the norm and relative phase of the wave function itself, on an individual system.  Since expectation values have physical meaning, the PBR theorem and protective measurements both imply that a quantum state represents physical reality.  Would it be right to say that protective measurements anticipate the PBR result?  In this section, we show that the answer to this question cannot be a simple Yes or No:  although close in spirit, protective measurements and the PBR theorem make different and complementary statements about the physical reality of quantum states.  First, however, we address the question of what it means to represent physical reality---a question that is not straightforward in quantum theory.

Hartle \cite{hartle} claims that the quantum state is not an objective property of the system, because no assertion about the state of the system ``can be verified by measurements on the individual system without knowledge of the system's previous history''.  Indeed, if we are given a single system in an unknown quantum state, protective measurements cannot identify its state, any more than other measurements can.  Hartle's conclusion is therefore that a quantum state is a property of an ensemble, but not a property of any individual system. (See also \cite{ball}.)  If so, then neither the PBR theorem not protective measurements make any statement about the reality of the quantum state of a single system.

Hartle's criterion---measurability without prior knowledge---is suitable for the classical world, but it rules out discussion of a single quantum system, and is thus unsuitable for the quantum world. It does not allow attribution of \textit{any} contingent property to individual quantum systems. The quantum world requires a more subtle criterion.

A better criterion for attributing a property to an individual quantum system is that of EPR (italics in the original):  ``{\it If, without in any way disturbing a system, we can predict with certainty ({i.e.} with probability equal to unity) the value of a physical quantity, then there exists an element of physical reality corresponding to this physical quantity.}"  We cannot apply this criterion in the case considered by EPR because of the failure of locality, but we can apply it here.  For example, if we have just found an isolated atom to be in an eigenstate of energy, we can be sure that a second measurement of its energy will yield the same result, and we can therefore attribute that energy to the atom.  Similarly, via protective measurements we can attribute to each quantum system a distinct reality defined by distinct properties.  Namely, protective measurements can be used to probe an individual system in the same way that other quantum measurements probe an ensemble.

How do we show that $\vert \Psi\rangle$ is a property of an ensemble?  We prepare an ensemble of systems in the state $\vert \Psi\rangle$, and perform measurements on that ensemble.  We are not totally ignorant of the preparation; on the contrary, we must know how the state is prepared in order to assign the properties we find to $\vert \Psi\rangle$.  If we were totally ignorant of the preparation, we couldn't even be sure of having an ensemble of identical systems.  We can view protective measurements in the same light, but instead of preparing an ensemble of many systems in the state $\vert \Psi\rangle$, we prepare a single system in the state $\vert \Psi\rangle$ and protect the state from changing during the (extended) measurement.  When we do so, and measure the expectation values of any operator we wish, we obtain values that define the quantum state uniquely.  For example, a measurement of a projection operator $\vert \Psi\rangle\langle \Psi \vert$ yields $1$ in the state $\vert \Psi\rangle$ and less than $1$ in any other state.  Since (protective) measurements on any given quantum state yield expectation values that differ from measurements on any distinct quantum state, each quantum state represents a distinct reality.

Thus protective measurements show the reality of a single system in a quantum state. But can they be used to prove that two quantum states inevitably represent different realities?  The answer to this question has to be negative, because if two nonorthogonal states $\vert\Psi_1\rangle$ and $\vert\Psi_2\rangle$ are possible inputs for a single measuring device, then the possible outputs cannot be orthogonal. Concretely, let us consider a measuring device prepared in a neutral state $\vert \chi_0 \rangle$ and coupled to either $\vert\Psi_1\rangle$ or $\vert\Psi_2\rangle$.  To discriminate unambiguously between these two states, the measuring device coupled to $\vert \Psi_i \rangle$ must evolve into the respective pointer state $\vert \chi_i \rangle$, where $\langle \chi_2 \vert \chi_1\rangle = 0$. But then the initially $nonorthogonal$ states $\vert\chi_0 \rangle \otimes \vert \Psi_1\rangle$ and $\vert\chi_0 \rangle \otimes \vert \Psi_2\rangle$ must evolve into the $orthogonal$ states $\vert\chi_1 \rangle \otimes \vert \Psi_1\rangle$ and $\vert\chi_2 \rangle \otimes \vert \Psi_2\rangle$, which is impossible with unitary time evolution.

Thus the description of a quantum state via protective measurements leaves an ambiguity, which we can summarize as follows.  Suppose we prepare an ensemble of systems in the state $\vert \Psi_1 \rangle$, protect the state $\vert \Psi_1 \rangle$, and measure a long list of observables.  We will, in each case, obtain the expectation values of those observables in the state $\vert \Psi_1 \rangle$.  However, if we prepare the ensemble of systems in the state $\vert \Psi_2 \rangle$ (neither identical nor orthogonal to $\vert \Psi_1 \rangle$) and protect the state $\vert \Psi_1 \rangle$, the protection will leave a fraction $\vert \langle \Psi_2 \vert \Psi_1 \rangle \vert^2$ of the systems in the state $\vert \Psi_1\rangle$.  Now if we measure the same long list of observables, we will obtain a sub-ensemble of systems yielding the expectation values of those observables in the state $\vert \Psi_1 \rangle$.  We then cannot eliminate the possibility that the reality $\lambda$ underlying some of the systems prepared in the state $\vert \Psi_2 \rangle$ is compatible with $\vert \Psi_1 \rangle$ as well as $\vert \Psi_2 \rangle$; we do not have a no-go theorem.    One could consider deriving such a theorem by considering, as PBR do, many systems prepared independently in either the state $\vert \Psi_1 \rangle$ or in the state $\vert \Psi_2 \rangle$, but such a derivation would have little to do with protective measurements.

This difference between protective measurements and the conclusion of PBR demonstrates the inherent inaccessibility of quantum reality.  We already knew that orthogonal states represent different physical realities.  We now know that nonorthogonal states, as well, represent different physical realities.  If they represent different physical realities, which observable can we measure to distinguish one physical reality from the other?  Of course, there is no such observable; if there were, two nonorthogonal states would be its nondegenerate eigenvectors.  Unitarity prevents us from distinguishing two nonorthogonal quantum states, and the PBR theorem implies that this indistinguishability does not arise because two quantum states can represent the same reality. They cannot.

To conclude, both protective measurements and PBR can be seen as partial answers to a single question:  What do two quantum systems, described by two quantum states in a common Hilbert space, have in common?  Protective measurements tell us that if the two quantum states are the same, the systems have a lot in common, namely the expectation values of all operators measurable on the systems.  The PBR theorem tells us that if the two quantum states are different, the systems are in physically different states.

\begin{acknowledgments}
We thank Prof. Lev Vaidman for Ref. \cite{cohen}.  This publication was made possible through the support of grants from the German-Israel Foundation (grant no. 1054/09), from the John Templeton Foundation (Project ID 43297), from the Israel Science Foundation (grant no. 1190/13).  The opinions expressed in this publication are those of the authors and do not necessarily reflect the views of any of these supporting foundations.

\end{acknowledgments}

\end{document}